\documentstyle[twoside,fleqn,espcrc2,epsf]{article}
\newcommand{\bee}{\begin{equation}}
\newcommand{\ee}{\end{equation}}
\newcommand{\beea}{\begin{eqnarray}}
\newcommand{\eea}{\end{eqnarray}}

\title{Topological structure in the SU(2) vacuum\thanks{Talk 
presented by T.G.\ Kov\'acs}\thanks{Research supported by
DOE grant DE--FG02--92ER--40672}}

\author{Thomas DeGrand, Anna Hasenfratz and
        Tam\'as G. Kov\'acs \\[2mm] 
        Department of Physics, 
        University of Colorado\\ Boulder, CO 80309-390, USA
       }

\begin{document}

\begin{abstract}
We study the topological content of the vacuum of
   $SU(2)$ pure gauge theory using lattice simulations. We
use a smoothing process based on the renormalization group equation.
This removes short distance fluctuations
but preserves long distance structure.  
The action of the smoothed configurations
 is dominated by instantons, but they still
show an area law for Wilson loops with an unchanged string tension.
The average radius of an instanton is about 0.2 fm, at a density of about 2 fm${}^{-4}$.
\end{abstract}

\maketitle

Based on phenomenological models, it has been argued that instantons 
are largely responsible for the low energy hadron and glueball spectrum 
\cite{Diakonov}. Instanton liquid
models attempt to reproduce the topological content of the QCD 
vacuum and conclude that hadronic correlators in the instanton liquid 
show all the important properties of the corresponding full
QCD correlators. These models appear to capture the essence of
the QCD vacuum, but their derivations involve a number of uncontrolled 
approximations and phenomenological parameters.

Lattice methods are the only ones we presently have, which might 
address this connection starting from first principles.
Lattice studies of instantons can suffer from several 
difficulties. An unambiguous topological charge can be assigned only to 
continuous gauge field configurations living on a continuum
space-time. On the lattice, the charge can only be defined as
that of an interpolated continuum filed configuration. This 
interpolation however is non-unique on
Monte Carlo generated lattice configurations. 

Another problem is connected to the fact that while the continuum
gauge field action is scale invariant, the lattice regularisation
breaks this invariance and the action of lattice instantons typically
depends on their size. This might distort the size distribution 
of instantons and in particular can lead to an overproduction of small 
instantons which can even spoil the scaling of the topological 
susceptibility.

The framework of classically perfect fixed point (FP) actions \cite{HN}
is particularly suitable to address these problems. Fixed point actions
can be shown to have scale invariant instanton solutions and there 
are no charge 1 objects with an action lower than the continuum instanton
action. The FP context also gives a consistent way of interpolation
to define the topological charge.

The fixed point action for any given configuration $V$
on a lattice with lattice spacing $a$ is defined by the weak coupling
saddle-point equation
\bee
S^{FP}(V)=\min_{ \{U\} } \left( S^{FP}(U) + T(U,V)\right),  
    \label{eq:saddle}
\ee
where $T$ is the blocking kernel of a real-space RG transformation
from the fine lattice (spacing $a/2$) to the coarse lattice $(a)$ and
the minimum is taken over all fine lattice configurations $U$. The minimising
fine configuration $U_{\mbox{\scriptsize min}}$ is the smoothest possible
of those $U$'s that block into $V$. It is a very 
special interpolating configuration which, in the weak 
coupling limit, gives the largest contribution to the path integral defining
the RG transformation for the given coarse configuration $V$. Finding
$U_{\mbox{\scriptsize min}}$ for a given $V$ will be referred to as
``inverse blocking''. In principle inverse blocking can be repeated
several times until the resulting configuration on the finest grid becomes smooth
enough that any ``sensible'' definition of the topological charge gives
the same integer value. 

In the remainder of the paper we discuss the implementation of the above
ideas for the 4d SU(2) gauge theory. For more details we refer the reader
to \cite{TAT}. Another implementation can be found in Ref.\ \cite{MMP}.
Our calculations were performed at several lattice
spacings between 0.1-0.18 fm with an approximate FP action.
At these values of the lattice spacing the total charge is well defined already after
one step of inverse blocking but individual instantons cannot be identified
at this stage. Further iteration of the inverse blocking is presently
impossible due to computer memory limitations. Therefore
we used a smoothing cycle based on inverse blocking followed by
a blocking step but on a different coarse sublattice, diagonally shifted by $a/2$.

In order to understand how this works we note that, although the inverse blocked
lattice has a physical lattice spacing $a/2$ as measured by any long-distance
observable, locally it is much smoother than typical Monte Carlo 
generated lattice configurations with the same lattice spacing. Nevertheless
this locally very smooth configuration, by construction, still blocks back
into the given $V$. This is ensured by the delicate coherence present
in the $U_{\mbox{\scriptsize min}}$ configuration on a distance scale of $a$.
The diagonal shift by $a/2$ before blocking destroys exactly this coherence
and as a result the shifted $U_{\mbox{\scriptsize min}}$ blocks into
a configuration that is locally much smoother than $V$ was. 
\begin{figure}[htb]
\begin{minipage}{7.0cm}
\epsfxsize=7.0cm \epsfysize=7.0cm
\epsfbox{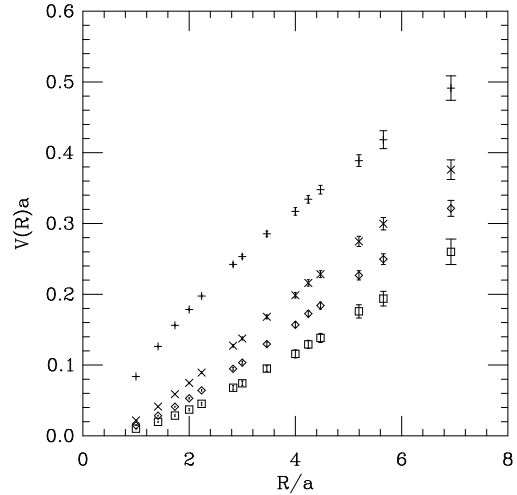}
\end{minipage}
\vspace{-0.7cm}
\caption{ Potential measured on the inverse blocked $16^4$ $\beta=1.5$
configurations after performing different numbers of smoothing cycles
(plusses 1 cycle, crosses 3 cycles, diamonds 5 cycles and squares 9 cycles).}
\label{fig:pot_cycle}
\end{figure}
                                  
After one such cycling step the shortest distance
fluctuations are considerably reduced but long-distance features are preserved.
 Since the lattice size does not change,
this step can be repeated several times. The stability of the long 
distance physical properties of the configurations can be demonstrated
by the invariance of the string tension (see Fig.\ref{fig:pot_cycle}). 
The total topological charge --- as measured in each step on the fine grid ---
as well as artificially laid down instantons and I-A pairs also turned out 
to be unchanged by cycling. After about 6 cycles the locations and sizes 
of individual instantons could be identified.

While their locations were quite stable from 
the stage where we could  reliably identify
them, the size of some instantons kept changing
slowly throughout the iteration. This was taken into account by extrapolation
when measuring sizes.
This might explain why on one of our
configurations Ref.\ \cite{Overlap} found that the 
topological charge --- as measured with the fermionic overlap ---
changed over cycling. It is very likely that
over repeated cycling an already existing small instanton grew above
the threshold below which the overlap cannot detect small instantons.

\begin{figure}[tb]
\begin{minipage}{7.0cm}
\epsfxsize=7cm \epsfysize=7cm
\epsfbox{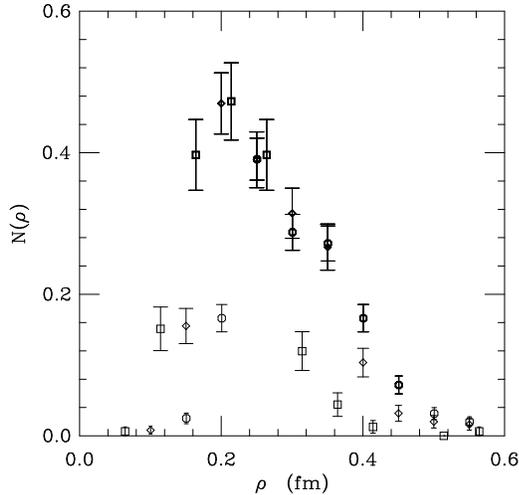}
\end{minipage}
\vspace{-0.7cm}
\caption{ The density distribution of instantons. 
Data at $a=0.116(2)$ fm are given by squares, $a=0.144(1)$ fm diamonds,
and $a=0.188(3)$ fm octagons. The bold data points are ones for which the instanton
radius is large compared to the lattice spacing and small compared to the
simulation volume.}
\label{fig:density}
\end{figure}

The instanton size distribution obtained at different values of the lattice
spacing --- as defined by the Sommer parameter --- is shown 
in Fig.\ \ref{fig:density}. We obtain a topological susceptibility
\bee
  \chi_t = \frac{\langle Q^2 \rangle}{V} = (230(10) MeV)^4,
\ee
which is about 20\% larger than the values obtained with improved 
cooling \cite{Forcrand} and the heating method 
\cite{DiGiacomo}.
                                           
\begin{figure}[tb]
\begin{minipage}{7.0cm}
\epsfxsize=7.0cm \epsfysize=7.0cm
\epsfbox{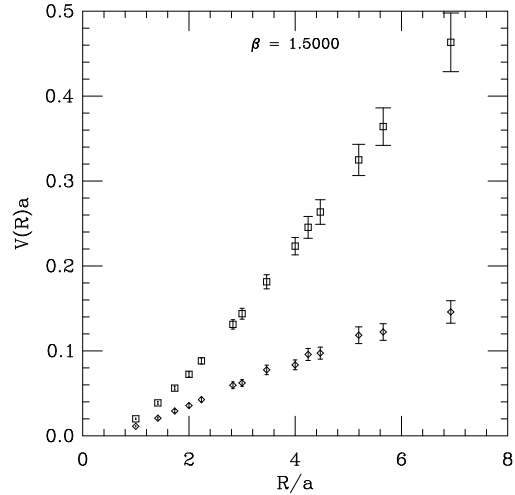}
\end{minipage}
\vspace{-0.7cm}
\caption{The heavy quark potential measured on the artificially 
produced instanton configurations (diamonds) and the potential measured on
the corresponding 9 times smoothed ``real'' configurations (squares),
at $\beta=1.5$.}
\label{fig:inst_pot}
\end{figure}

The smoothed configurations have essentially the same long-distance physical
properties as the unsmoothed ones. On the other hand about 70\% of
their action can be accounted for by the instantons. This alone
suggests that instantons might explain most of the long-distance features 
of QCD. The most straightforward way to check this is to
prepare artificial configurations by laying down instantons in 
exactly the same way as they were found on the smoothed configurations,
and then to compare the physical properties of these artificial 
configurations with the real smoothed ones.
As an illustration in Fig.\ \ref{fig:inst_pot} we show 
the heavy quark potential measured with timelike Wilson loops. 
The comparison shows that instantons are not very likely to be responsible
for confinement. Since we have no information
about the relative orientation of instantons in group space, in the artificial
configurations we just put them aligned. Work is in progress to
study the case of randomly oriented instantons.  

\newcommand{\PL}[3]{{Phys. Lett.} {\bf #1} {(19#2)} #3}
\newcommand{\PR}[3]{{Phys. Rev.} {\bf #1} {(19#2)}  #3}
\newcommand{\NP}[3]{{Nucl. Phys.} {\bf #1} {(19#2)} #3}
\newcommand{\PRL}[3]{{Phys. Rev. Lett.} {\bf #1} {(19#2)} #3}
\newcommand{\PREPC}[3]{{Phys. Rep.} {\bf #1} {(19#2)}  #3}
\newcommand{\ZPHYS}[3]{{Z. Phys.} {\bf #1} {(19#2)} #3}
\newcommand{\ANN}[3]{{Ann. Phys. (N.Y.)} {\bf #1} {(19#2)} #3}
\newcommand{\HELV}[3]{{Helv. Phys. Acta} {\bf #1} {(19#2)} #3}
\newcommand{\NC}[3]{{Nuovo Cim.} {\bf #1} {(19#2)} #3}
\newcommand{\CMP}[3]{{Comm. Math. Phys.} {\bf #1} {(19#2)} #3}
\newcommand{\REVMP}[3]{{Rev. Mod. Phys.} {\bf #1} {(19#2)} #3}
\newcommand{\ADD}[3]{{\hspace{.1truecm}}{\bf #1} {(19#2)} #3}
\newcommand{\PA}[3] {{Physica} {\bf #1} {(19#2)} #3}
\newcommand{\JE}[3] {{JETP} {\bf #1} {(19#2)} #3}
\newcommand{\FS}[3] {{Nucl. Phys.} {\bf #1}{[FS#2]} {(19#2)} #3}



\begin{thebibliography}{99}


\bibitem{Diakonov}
D.~Diakanov, hep-ph/9602375;
T.~Sch\"afer and E.~V.~Shuryak, hep-ph/9610451.

\bibitem{HN}
P.~Hasenfratz and F.~Niedermayer, Nucl. Phys. B414 (1994) 785;
F.~Niedermayer, Nucl.\ Phys.\ B (Proc.\ Suppl.) {\bf 53} (1997) 56;
P.~Hasenfratz, talk at this conference.

\bibitem{TAT}
T.~DeGrand, A.~Hasenfratz and T.G.~Kov\'acs, hep-lat/9705009.


\bibitem{MMP}
M.~Feurstein, E.-M.~Ilgenfritz, 
M.~M\"uller-Preussker and S.~Thurner, Berlin  preprint HUB-EP-96/59,
hep-lat/9611024 and talk at this conference.

\bibitem{Overlap} 
R.~Narayanan and R.L.~Singleton, hep-lat/9709014.

\bibitem{Forcrand} 
P.~de~Forcrand, M.~ Garcia~Perez and I.O.~Stamatescu, hep-lat/9701012 and
talk at this conference.

\bibitem{DiGiacomo}
B.~Alles, M.~D'Elia, A.~Di Giacomo, hep-lat/9706016 and talk at this
conference.


\end{thebibliography}
\end{document}